\documentstyle[twocolumn,prl,aps]{revtex}
\begin{document}
\draft
\title{Suppression of antiferromagnetic correlations by dipole-type impurities
in lamellar cuprates}
\author{I. Ya. Korenblit, V. Cherepanov, Amnon Aharony, and O. Entin-Wohlman}
\address{School of Physics and Astronomy, Raymond and Beverly Sackler Faculty of
 Exact Sciences,\\
Tel Aviv University, Tel Aviv 69978, Israel}
\date{\today}
\maketitle
\begin{abstract}
In doped lamellar cuprates, 
localized holes create ferromagnetic bonds and cause spin canting
similar to that caused by magnetic dipoles. At low temperatures, these
dipoles are frozen, behaving like quenched correlated random fields.
Renormalization group methods are used to show that such impurities
cause a strong reduction of the two-dimensional antiferromagnetic correlations 
in the non-linear $\sigma$ model, and a related decrease in the 
three dimensional N\'eel temperature, in quantitative agreement with
experiments.
\end{abstract}

\narrowtext
Many lamellar copper oxides exhibit strong two dimensional (2D)
Heisenberg 
antiferromagnetic (AFM) correlations, with a correlation length $\xi_{2D}$,
and a 3D AFM ordering, at a N\'eel temperature $T_N$.
Doping, which
is believed to introduce a concentration $x$ of
electronic holes onto the oxygen ions in the CuO$_2$
planes, leads to a rapid decrease in both $\xi_{2D}$ and $T_N$, and to
a disappearance of the AFM order above $x_c \approx 2\%$
This decrease has been attributed
to frustration, due to strong ferromagnetic (FM) exchange on the
Cu-O-Cu bonds which have localized holes
 \cite{aharony}.
The FM bonds cause canting of the AFM moments, which is described at long
distance as resulting from effective dipoles \cite{aharony,villain}.
This frustration was also predicted to yield a magnetic spin glass phase
for $x>x_c$, as recently confirmed in detail in 
doped La$_2$CuO$_4$ ~\cite{choua}.

At finite temperature $T$, the 2D AFM correlations of the undoped systems are
excellently described by the classical non-linear-$\sigma$-model (NL$\sigma$M)
with a renormalized stiffness constant $\rho_s$~\cite{CHN}.
A few years ago, Glazman and Ioselevich (GI) \cite{glazman} added dipolar
defects to this model, integrated them out of
the partition function and described their
effects on
$\xi_{2D}$ as a renormalization of $\rho_s$.
Although the locations ${\bf r}_\ell$ of
the dipole-like impurities and the unit vectors along the corresponding
Cu-O-Cu bonds, ${\bf a(r}_\ell)$, 
are randomly {\it quenched}, each impurity involves
an
effective dipole moment $M{\bf m(r}_\ell)$ (with magnitude $M$)
which is still free to reach
{\it annealed}
equilibrium in the presence of all the other dipoles. GI used an annealed
averaging, expanded the renormalized
$\rho_s$ to leading order in $x\rho_s/T$ and
used qualitative arguments to deduce a reentrant phase diagram for $T_N(x)$.
Here we argue
that
at sufficiently low $T$ the dipole moments (which interact via randomly quenched
dipole-dipole interactions) freeze in a random spin glassy way on the
relevant length scales \cite{hertz}.
Therefore, they should also be treated as {\it quenched} variables. We thus
solve the classical NL$\sigma$M in the presence of such quenched
random perturbations,
using the renormalization group (RG), and obtain $\xi_{2D}$ for {\it all} 
the relevant values of $T$ and $x$.
In principle, one should use the {\it annealed} averaging for the first few RG
iterations, until the renormalized cell contains more than a few impurities,
and then (at low $T$) switch to the quenched averaging. 
To leading order in $x$, the two types of average give the same results,
and our results agree with GI. However, at larger $x$ our results differ
qualitatively from those of GI, and agree with experiments. 

Specifically,
we express our results in terms of the renormalized parameters
$t=T/\rho_s$ and $\lambda=Ax$, where $A$ is a dimensionless number.
For $t<\lambda$ we find that $\xi_{2D}(t,\lambda)$
is independent of $t$, and is given by
\begin{equation}
\xi_{2D}(t,\lambda) \approx C(0,\lambda)\exp(2\pi/3\lambda).
\label{III9}
\end{equation}
where $C(0,\lambda)\approx C_0 \lambda^\omega$. Fig. 1 shows this expression,
with $C_0=2.8\AA$, $\omega=0.8$ and $A=20$, in comparison with 
experimental\cite{keimer,hayden} 
and numerical\cite{GM} results.
  
At higher $T$, we find
\begin{equation}
\xi_{2D}(t,\lambda)\approx C(t,\lambda) \exp\Bigl(\frac{2\pi}{t}\Bigl[1-
\frac{\lambda}{t}
+\frac{\lambda^2}{3t^2}\Bigr]\Bigr),\label{III8}
\end{equation}
with a slowly varying $C(t,\lambda)$. This
agrees with GI to leading order in $\lambda/t$, and its
exponential part matches continuously
with Eq. (\ref{III9}) at $t=\lambda$.
Fig. 2 shows this expression, using the known undoped values
$C(t,\lambda)=1.92\AA$ and
$2\pi\rho_s=150$meV from Ref. \onlinecite{keimer}, 
for $x=0.0225, 0.029$ and 0.036, compared
with experimental data for $x=0.020(5), 0.030(5)$ and 0.040(5)~\cite{keimer}. 
The figure
also shows our theoretical predictions for $t<\lambda$, from Fig. 1 (the
use of two different prefactors causes a discontinuity around $t=\lambda$,
smoothed by the dotted lines).
The agreement with the data is comparable with (if not better than) that
of the approximate expression
$\xi^{-1}(T,x) =\xi^{-1}(0,x) + \xi^{-1}(T,0)$ used in Ref. \onlinecite{keimer}
with no theoretical basis (shown by the dashed lines).

The 3D N\'eel temperature $T_N(x)$
is roughly
given by the solution of $\alpha \xi_{2D}^2=1$, where $\alpha$ represents
either the relative
interplane exchange, $J_{\perp}/J \sim J_\perp/2\pi\rho_s$, or the in-plane
spin relative anisotropy.
Using $\alpha=10^{-4}$ for La$_2$CuO$_4$\cite{keimer}, 
Eq. (\ref{III9}) now yields
$x_c=0.0183$, in good
agreement with the experimental value $x_c=0.0175$ from
Ref. \onlinecite{shirane} (but in conflict with $x_c=0.027$ from
Ref. \onlinecite{kasegava}).
Approximating 
$C(t,\lambda)$ by the constant $C(0,\lambda_c) = C(0,0.366) =
1.26 \AA$, the above expressions yield $T_N(x)$ as shown in Fig. 3. Using
the two different prefactors mentioned after Eqs. (\ref{III9}) and
(\ref{III8}) would imply changes in $T_N$ by less
than $10\%$.
Given that all the parameters were already determined from the experiments
on $\xi_{2D}$, our $T_N(x)$ is in very good agreement with the measurements
of Refs. \onlinecite{saylor,cho}, based on Sr-doped La$_2$CuO$_4$. The
disagreement with the data of Ref. \onlinecite{chen} may be due to possible
systematic errors in $x$
(these data were on O-doped samples, with $x$ determined by Hall
measurements).
In contrast with GI's heuristic arguments, which led to a reentrance, our
theory predicts that the line $T_N(x)$ remains vertical at $x=x_c$
up to the value
where $t_N=\lambda_c\equiv Ax_c$.

The conventional method to treat
the NL$\sigma$M expands the order parameter unit vector {\bf n(r)} 
of antiferromagnetism about a
spatially uniform ordered state \cite{brezin,nelson}. However, since the dipole-doped
system has no AFM long range order in 2D
even at
$T=0$, we preferred the original RG approach
by Polyakov \cite{polyakov1}, which preserves rotational symmetry
at all steps of the RG procedure. In fact, the conventional method
generated local uncorrelated random fields which should not be there\cite{else}.

Following GI, we describe the system
by the reduced Hamiltonian
\begin{equation}
{\cal H}={\cal H}_{\rm pure}+{\cal H}_{\rm int},\label{II1}
\end{equation}
where ${\cal H}_{\rm pure}$ is the classical NL$\sigma$M for the undoped
system, representing
the long wave length Hamiltonian
related to the fluctuations of ${\bf n}({\bf r})$, 
\begin{equation}
{\cal H}_{\rm pure}=\frac{1}{2t}\int d{\bf r}\sum_{i,\mu}
(\partial_i n_{\mu})^{2},\label{II2}
\end{equation}
where $i=1,...,d$ and $\mu =1,..,{\cal N}$ run over
 the spatial cartesian components
and over the spin components, respectively, and
$\partial_i \equiv \partial/\partial x_i$.
${\cal H}_{\rm int}$ was constructed by GI
to reproduce the dipolar effects discussed
in Ref. \onlinecite{aharony},
\begin{equation}
{\cal H}_{\rm int}=\frac{1}{t}\int d{\bf r}\sum_{i}{\bf f}_{i}({\bf r})
\cdot \partial_i {\bf n},\label{II3}
\end{equation}
with
\begin{equation}
{\bf f}_{i}({\bf r})=M\sum_{\ell}\delta({\bf r}-{\bf r}_{\ell})
a_{i}({\bf r}_{\ell}){\bf m}({\bf r}_{\ell}).\label{II4}
\end{equation}

We now decompose
${\bf n}({\bf r})$ into a slowly varying part, given by
the unit vector $\tilde{{\bf n}}({\bf r})$, 
and $({\cal N}-1)$ fast variables $\phi_{\mu}({\bf r})$
such that
\begin{equation}
{\bf n}({\bf r})=\tilde{{\bf n}}({\bf r})
\sqrt{1-\phi^{2}({\bf r})}+\sum_{\mu =1}^{{\cal N}-1}\phi_{\mu}({\bf r})
{\bf e}_{\mu}({\bf r}),\label{II5}
\end{equation}
with
$\phi^{2}({\bf r})=\Sigma_{\mu =1}^{{\cal N}-1}\phi_{\mu}^{2}({\bf r})$.
The unit
vectors $\tilde{{\bf n}}({\bf r})$ and ${\bf e}_{\mu}({\bf r})$,
$\mu =1,...,{\cal N}-1, $ form an orthonormal basis. The Fourier transform
of the fast fields $\phi_{\mu}$ is restricted to wave vectors
${\bf q}$ in the range $b^{-1}\leq q\leq 1$. The upper
 bound is the inverse of
the microscopic length, and $b=e^{\ell}$ is the length rescale factor
for the renormalization procedure.
These $q$ values are to be integrated out.
After the iteration, the correlation length
$\xi$ is renormalized into $\xi/b$.

The next stage involves integration over the fast variables
 $\phi_\mu$ \cite{polyakov1}. This yields, 
like in Ref.~\onlinecite{glazman}, an effective 2D dipole-dipole-type 
interaction 
between dipoles separated by a distance $r<b$.
The  interaction decays as $r^{-d}$.
This renormalization can be repeated up to length scale $b=L_0$, and 
the resulting
dipole-dipole interaction then applies to all the moments within
the renormalized cell of size $L_0$.
Clearly, to have a non-zero interaction within the cell we need
$L_0 > x^{-1/d}$, where 
 $x^{-1/d}$ is of the order of
the average distance between impurities.
We also require $L_0 \ll \xi_{2D}$, or else the impurities have no effect
on the ``pure" Heisenberg model behavior.

A central issue is the averaging over the random
fields 
${\bf f}_i$. This includes
averaging over the quenched random variables ${\bf r}_\ell$ and  
${\bf a(r}_\ell)$, and over
the variables ${\bf m(r}_\ell)$, which are coupled via the above dipole-dipole
interaction.
 We use $[a_i({\bf r}_1)a_j({\bf r}_2)]=
\delta_{ij}\delta({\bf r}_{12})x/d$, where $[...]$ denotes the quenched
average, ${\bf r}_{12}={\bf r}_1-{\bf r}_2$, and find
$[f_{i\mu}({\bf r}_1)f_{j\nu}({\bf r}_2)]=\delta_{ij}\delta({\bf r}_{12})
\Lambda_{\mu\nu}({\bf r}_1)$,
with
$\Lambda_{\mu\nu}({\bf r})=
m_\mu({\bf r})m_\nu({\bf r})
M^2 x/d$.
At temperatures higher than the dipole-dipole interaction at mean distances
(with energy $E_{\rm dd}$ of order $\lambda\rho_s$),
one can follow GI and treat the $m_\mu$'s as
{\it annealed}.
To leading order, this implies that
$\langle m_\mu({\bf r}_1)m_\nu({\bf r}_2)\rangle =
 \delta_{\mu\nu}\delta({\bf r}_{12})/{\cal N}$,
and thus
$\langle \Lambda_{\mu\nu} \rangle = \delta_{\mu\nu}\Lambda \equiv \delta_{\mu\nu}
M^2x/(d{\cal N})$.
Given this relation, one derives the same leading order recursion relation for 
$t$ as presented below for the quenched case, while 
$\Lambda$ remains unrenormalized in 2D~\cite{else}.

We now argue that  for sufficiently 
low temperatures one should treat the ${\bf m}$'s
as {\it quenched} variables.
It was argued (see Refs.~\onlinecite{hertz} and references therein) that
spin glasses  with long-range interactions, decaying 
as $r^{-d}$, are at their lower
critical dimension. Hence, the spin-glass correlation length,
$\xi_{sg}$, increases exponentially with decreasing $T/E_{\rm dd}$. 
Therefore, when $T \ll E_{\rm dd}$, we expect $\xi_{sg} \gg L_0$.
When the cell size $L_0$ is larger than $x^{-1/d}$, 
each cell contains many impurities, and
the strong spin-glass correlations will yield freezing of the
dipole
moments $\langle{\bf m}({\bf r}_{k})\rangle$ within each cell
into an apparent spin glass phase, with an
Edwards-Anderson order parameter \cite{hertz}
$Q \equiv [\langle m_\mu({\bf r}_{k})\rangle^2].$
At very low $T$, this becomes $Q \approx 1/{\cal N}$.
We thus end up with a renormalized lattice in which
the variables ${\bf m}({\bf r}_{k})$
are frozen in random directions, and all the factors 
in ${\bf f}_{i}$ become quenched.
Using the quenched average $[m_\mu({\bf r}_1)m_\nu({\bf r}_2)]=
\delta_{\mu\nu}\delta({\bf r}_{12})Q$ in $\Lambda_{\mu \nu}$,
we now have
\begin{equation}
\bigl[f_{i\mu}({\bf r}_1)f_{j\nu}({\bf r}_2)\bigr]=\lambda\delta_{\mu\nu}
\delta_{ij}\delta ({\bf r}_{12}),\label{II22}
\end{equation}
where
$\lambda =M^{2}Qx/d \approx \Lambda.$
Fourier transforming the variables in Eq. (\ref{II3}),
one can see that ${\cal H}_{\rm int}$ represents random fields with quenched
correlations of the form $[{\hat h}_\mu({\bf k}){\hat h}_\nu({\bf k}')]=
\lambda k^2 \delta_{\mu \nu} \delta({\bf k+k'})$, where $h_\mu=\Sigma_j\partial_jf_{j\mu}$.
Such correlations shift the lower critical dimension of the random field
Heisenberg problem from 4 down to 2~\cite{lacour,IM}, and at 2D one expects
Imry-Ma domains of sizes given by Eq. (\ref{III9}) above. Indeed,
 our detailed RG calculations confirm this expectation.

Integrating out the fast variables $\phi$, and rescaling the momenta,
yields the one-loop recursion relations \cite{else}
\begin{eqnarray}
\frac{dt}{d\ell}=-\epsilon t + \frac{{\cal N}-2}{2 \pi}t^2+\frac{{\cal N}-1}
{2\pi}t \lambda,\nonumber\\
\frac{d\lambda}{d\ell}=-\epsilon \lambda +\frac{{\cal N}-3}{2\pi}\lambda t
+\frac{{\cal N}}{2\pi}\lambda^2\label{lam}
\end{eqnarray}
to leading order in $\epsilon=d-2$.
The renormalization procedure also generates many new random terms,
which were not included in our initial ${\cal H}$.
However, all of these are irrelevant in the RG sense \cite{else}.

We now proceed  to calculate $\xi_{2D}$.
The prefacing annealed iterations up to length scale
$L_0=e^{\ell_0}$ are only expected to change $t$ and $\lambda$ by
factors of order unity. Thus, for $\ell > \ell_0$ we solve Eqs. (\ref{lam}) with
the approximate initial values
$t(\ell_0) \equiv t_0 = t + O(t^2,\lambda)
\approx t$ and $\lambda(\ell_0)\equiv \lambda_0= Ax$.
The solution is particularly simple for the
Heisenberg system, ${\cal N}=3$. At 2D one finds
\begin{eqnarray}
\lambda(\ell) &=&\lambda_{0}\Bigl(1-\frac{3\lambda_{0}}{2\pi}(\ell-\ell_0) \Bigr
)^{-1},\nonumber\\
t(\ell)&=&\lambda(\ell)\Bigl[1+\Bigl(\frac{\lambda_{0}}{t_{0}}-1\Bigr)
\Bigl(\frac{\lambda(\ell)}{\lambda_{0}}\Bigr)^{1/3}\Bigr]^{-1}. \label{III1}
\end{eqnarray}
There is  only one
fixed point, at $[\lambda ,t]=[0,0]$, and both quantities $t(\ell )$ and
$\lambda (\ell )$ flow away from it as $\ell $ increases. Using the 
standard scaling
relation
\begin{equation}
\xi(t,\lambda)=e^{\ell }\xi(t(\ell ),\lambda (\ell )),\label{III2}
\end{equation}
we can find $\xi $ from the matching condition
\begin{equation}
{\rm max}\bigl(t(\ell^{\ast}),\lambda(\ell^{\ast})\bigr)=2\pi  ,\label{III
3}
\end{equation}
where $\xi (t(\ell^{\ast}),\lambda (\ell^{\ast}))\equiv {\tilde C}(t,\lambda)$ 
is a
slowly
varying function of $t$ and $\lambda$.
For our 2D solution we find 
\begin{equation}
\xi_{2D}(t,\lambda)=L_0 {\tilde C}(t,x) 
\exp\Bigl(\frac{2\pi}{3 \lambda_0}(1-z^3)\Bigr),
\label{xixi}
\end{equation}
where $z^3=\lambda_0/(2 \pi)$ if $\lambda_0>t_0$ (i.e.
$\lambda(\ell^\ast) = 2 \pi$),
and $z$ is the solution
of the equation
\begin{equation}
z^3+(\lambda_{0}/t_{0}-1)z^2
=\lambda_{0}/2\pi\label{III4}
\end{equation}
if $\lambda_{0}<t_{0}$ ($t(\ell^\ast)=2 \pi)$.
In the former case, this yields the exponential part in Eq.   
(\ref{III9}) (where we have dropped the subscript in $\lambda_0$). From
Eq. (\ref{xixi}), $C(0,x)$ contains a factor $L_0 \sim x^{-1/d}$.
Another $x$-dependent factor will arise from higher order loops\cite{else}.
Thus, $C(0,x) \approx C_0 \lambda^\omega$, with $\omega$ of order unity
(to be calculated or fitted).

When $\lambda_0 < t_0$, so that $z$ is not too small,
one has $z \approx 1-\lambda_0/t_0$, and thus
our Eq. (\ref{xixi}) reduces to Eq. 
(\ref{III8}),
where $C(t,x)$ absorbs all the slowly varying additional factors.
In this limit, the prefactor $C(t,x) \approx C(t,0)$ is known from
the two-loop\cite{CHN} and three-loop\cite{HN} calculations.
For La$_2$CuO$_4$, $C(t,0)=1.92 \AA$.

Finally, we discuss the derivation of the 3D $T_N(x)$ in Fig. 3,
as deduced from
$\alpha \xi_{2D}^2 \sim 1$.
The critical value  $x_c$ (above which there is
no AFM long range order at any temperature) is given by
the solution of
\begin{equation}
\lambda_c=-4\pi/[3 \ln (C(0,\lambda_c)^2 \alpha)], \label{xc}
\end{equation}
with $\lambda_c=Ax_c$.
In fact, Eq. (\ref{III9}) is expected to give a good approximation for
$\xi_{2D}$ for a range of temperatures obeying $\lambda_0 > t_0$.
Therefore, the critical line $t_N(x)$ is expected to be practically
vertical when $t_N(x) < Ax$.

For small $x$, i.e. $\lambda_0 \ll t_0$, we can use Eq. (\ref{III8}),
and obtain
the linear approximation
\begin{eqnarray}
t_N(x)/t_N(0) \approx 
1 - Ax /t_N(0),\label{TN}
\end{eqnarray}
where  $t_N(0)$, the N\'eel temperature of the pure AFM, is the solution of
$t_N(0)=4\pi/\ln[1/\alpha C(t_N(0),0)^2]$.

For intermediate values of $x$ (below $x_c$)  Eq. (\ref{III8}) yields
\begin{equation}
t_N(x) = \lambda\left\{1 -\left[1 - \frac{3 \lambda}{4 \pi}
\ln\left({1\over  \alpha  C(t_N(x),x)^2}\right)\right]^{1/3}\right\}^{-1}.
\label{tm}
\end{equation}

The parameters used in Figs. 1-3 were chosen as follows:
The linear dependence of $(t/2\pi)\ln (\xi/C)$ on $1/t$ for   
a sample of
La$_2$CuO$_{4+\delta}$, with $T_N= 90$K~\cite{keimer}, gives
$\lambda=0.29$ ~\cite{korn}. From
Refs. \onlinecite{keimer,chen} we deduce that this value of $T_N$
corresponds to $x \approx 0.0145$, hence $A \approx 20$.
This is the only free parameter used to describe both $\xi_{2D}$ and
$T_N(x)$ for $t>\lambda$! 
The values of $C_0$ and $\omega$ were chosen to fit the data in Fig. 1.

At temperatures lower than $\sim 200 - 250$K, $\xi_{2D}$
does not depend on the temperature up to exponentially small
terms, of the order of $\xi(0,T)^{-1}$ ~\cite{keimer}. This nontrivial
property is reproduced well by the theory. At the highest
temperatures  the calculated values of
$\xi_{2D}$ are smaller than the measured ones. Perhaps, at such high
temperatures thermal fluctuations, which should decrease
the dipole moment, become important. In fact, at these temperatures one
might have to allow some annealed averaging, both on the dipole moments
and on their locations.

In conclusion, we employed the NL$\sigma$M to treat the  effect
of hole doping on the properties of lamellar copper oxides.
Our theory  seems to agree with experiments on both $\xi_{2D}(t,\lambda)$
 and $T_N(x)$. 

We have benefitted from many discussions with R.~J.~Birgeneau, 
A.~B.~Harris, A.~S.~Ioselevich, M.~A.~Kastner,
D.~E.~Khmelnitzkii, V.~V.~Lebedev, T.~Nattermann and M.~Schwartz. This
project has been supported by a grant from the U. S. - Israel Science
Foundation (BSF).

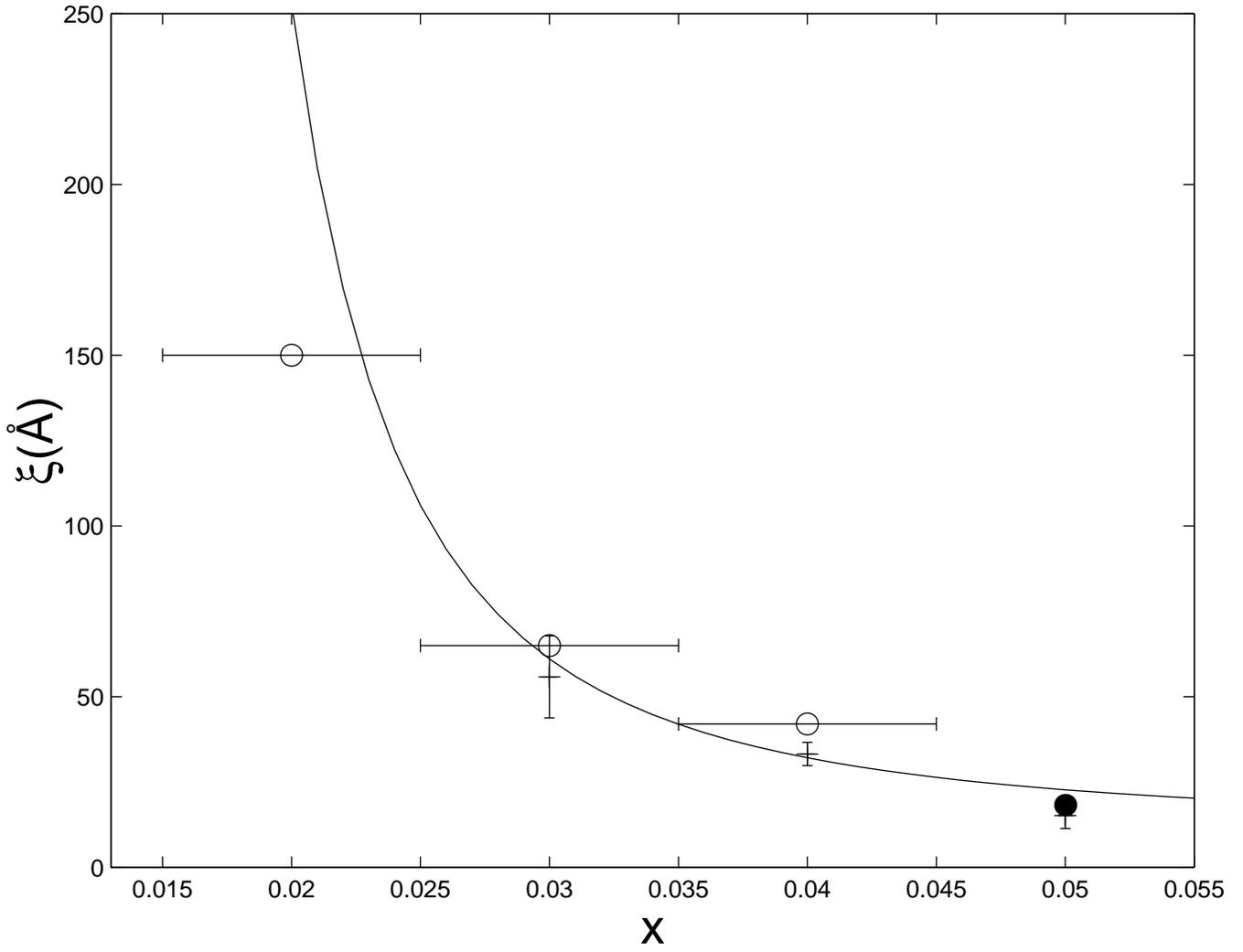
\begin{figure}
\caption{Dependence of $\xi_{2D}$ at $T=0$ on $x$.  The empty[2] and full[13]
 circles indicate
experiments, +'s show numerical simulation[14] data. The solid line 
represents Eq. (1)
with $C(0,\lambda)= 2.8 \lambda^{0.8} \AA$ and $\lambda=20x$.}

\end{figure}
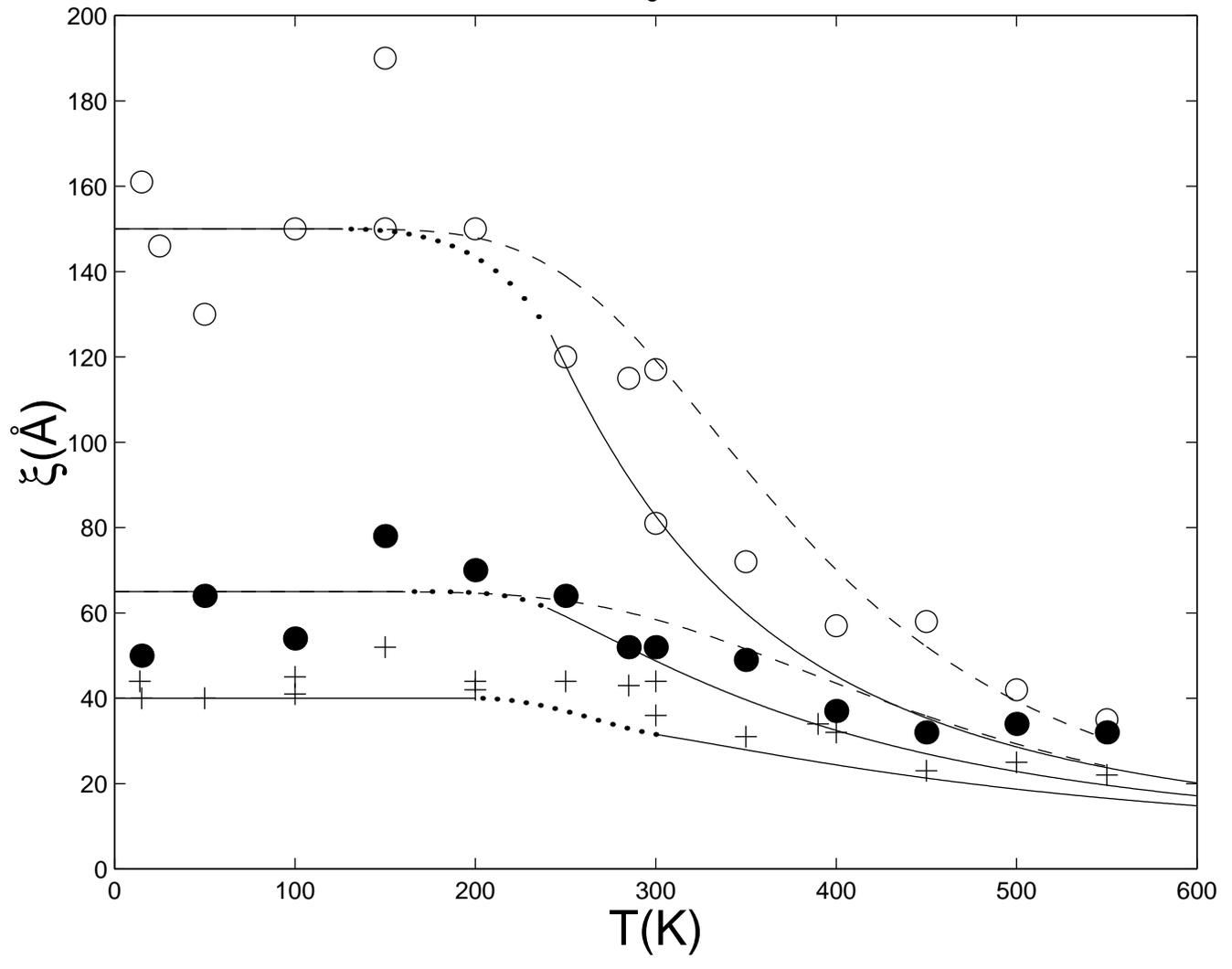
\begin{figure}
\caption{Dependence of $\xi_{2D}(t,\lambda)$ on $T$ for several concentrations.
Symbols are from experiments: empty circles for $x=0.02$, full circles for
$x=0.03$, +'s for $x=0.04$, all from Ref. 2. 
Full lines show Eqs. (1) (with $C=2.8\lambda^{0.8}\AA$) and (2) (with
$C=1.92\AA$), and dotted lines interpolate between
these low and high temperature theories.}
\end{figure}
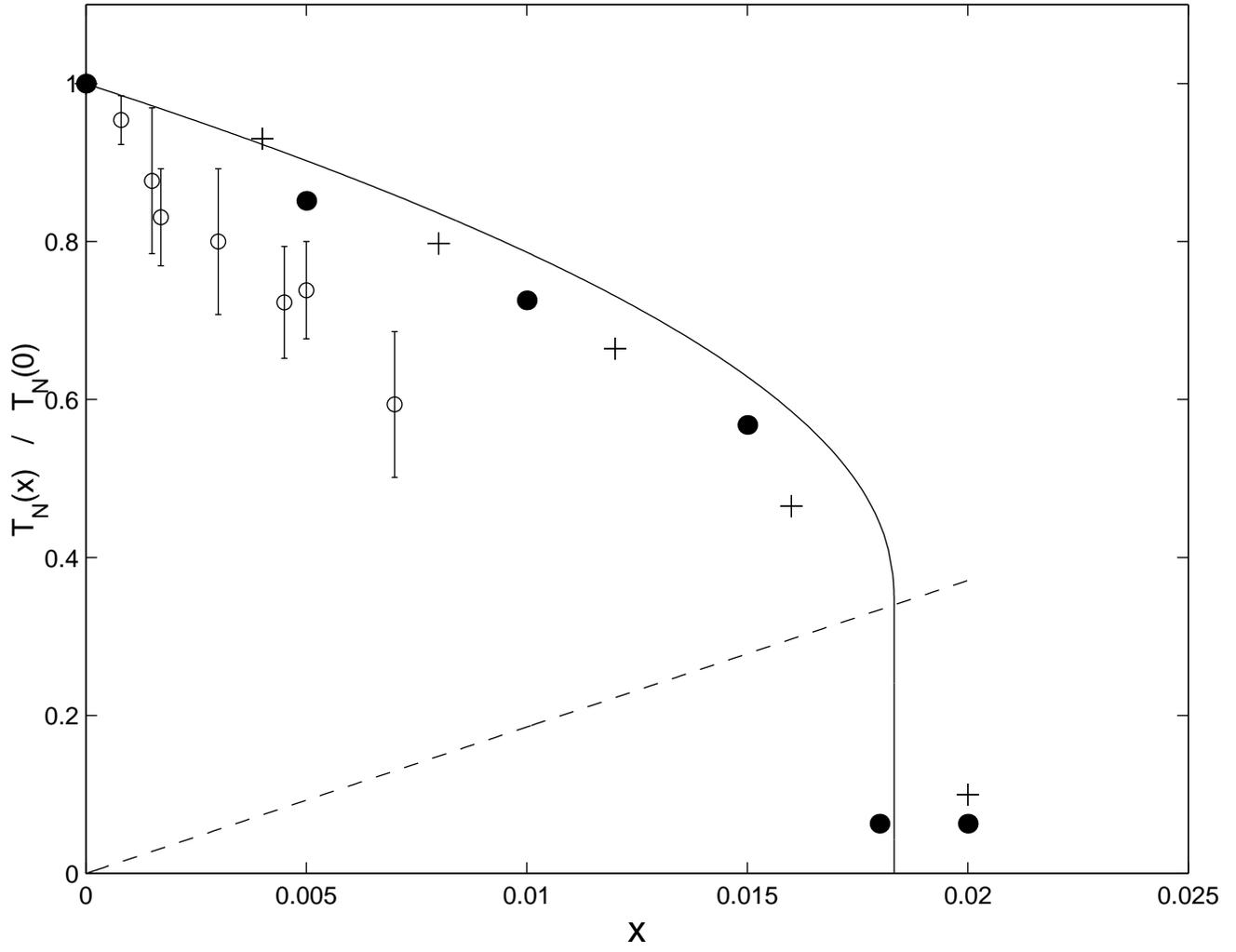
\begin{figure}
\caption{$T_N(x)/T_N(0)$ versus $x$. Full line is theory, and the points
are from experiments:  
 full circles
from Ref.~4 , +'s from Ref.~5, empty circles from Ref. 15.
Dashed line indicates $\lambda_0=t_0$.}
\end{figure}


\begin{references}
\bibitem{else} More details will be published elsewhere.
\end{references}
\end{document}